\documentclass[12pt]{amsart}

\usepackage{amssymb}
\usepackage{amsmath}
\usepackage{amscd}
\usepackage{latexsym}

\newtheorem{thm}{Theorem}[section]
\newtheorem{lemma}[thm]{Lemma}
\newtheorem{defn}[thm]{Definition}
\newtheorem{rmk}[thm]{Remark}
\newtheorem{coro}[thm]{Corollary}
\newtheorem{examp}[thm]{Example}
\newtheorem{prop}[thm]{Proposition}

\newtheorem{problem}{Problem}

\newcommand{\MO}{\mathord{\mathrm{MO}}}
\newcommand{\Sub}{\mathord{\mathrm{Sub}}}
\newcommand{\BSub}{\mathord{\mathrm{BSub}}}
\newcommand{\Image}{\mathord{\mathrm{Im}}}
\newcommand{\modif}{_0}

\begin{document}

\title{Subalgebras of Orthomodular Lattices}
\author{{\bf John Harding}%
\\ \it Department of Mathematical Sciences
\\ \it New Mexico State University, Las Cruces, NM 88003, USA
\\ \tt jharding@nmsu.edu
\\\and 
\\ {\bf Mirko Navara}%
\\ \it Center for Machine Perception, Department of Cybernetics
\\ \it Faculty of Electrical Engineering, Czech Technical University in Prague
\\ \it Technick\'a~2, 166~27 Praha, Czech Republic
\\ \tt navara@cmp.felk.cvut.cz}
\date{}


\maketitle
\markboth{John Harding and Mirko Navara}{Subalgebras of Orthomodular Lattices}

\begin{abstract}
Sachs \cite{Sachs} showed that a Boolean algebra is determined by its lattice of subalgebras. We establish the corresponding result for orthomodular lattices. We show that an orthomodular lattice $L$ is determined by its lattice of subalgebras $\Sub(L)$, as well as by its poset of Boolean subalgebras $\BSub(L)$. The domain $\BSub(L)$ has recently found use in an approach to the foundations of quantum mechanics initiated by Butterfield and Isham \cite{Isham1,Isham2}, at least in the case where $L$ is the orthomodular lattice of projections of a Hilbert space, or von Neumann algebra. The results here may add some additional perspective to this line of work. 
\end{abstract}

\section{Introduction}

In \cite{BLT} Birkhoff showed that the subalgebra lattice $\Sub(B)$ of a finite Boolean algebra $B$ with $n$ atoms is dually isomorphic to the lattice $P_n$ of partitions of an $n$-element set $\{1,\ldots,n\}$. Indeed, viewing $B$ as the power set of $\{1,\ldots,n\}$, the atoms of a subalgebra $S$ of $B$ provide a partition of $\{1,\ldots,n\}$, and each such partition arises this way. 

Sachs \cite{Sachs} showed every Boolean algebra $B$ is determined by its lattice of subalgebras. In fact, he showed that if $B,C$ are Boolean algebras and $\varphi\colon \Sub(B)\to \Sub(C)$ is a lattice isomorphism, then there is a Boolean algebra isomorphism $\varphi^*\colon B\to C$ with $\varphi(S)=\varphi^*[S]$ for each subalgebra $S$ of $B$.
Here $\varphi^*[S]$ denotes the image $\{\varphi^*(s): s\in S\}$ of the set $S$ under the map~$\varphi^*$.
Further, this map $\varphi^*$ is unique provided that $B$ has other than four elements. 
Gr\"{a}tzer {\em et. al. } \cite{Gratzer} characterized the lattices arising as $\Sub(B)$ for some Boolean algebra $B$ as those algebraic lattices where the interval $[0,k]$ beneath each compact element $k$ is dually isomorphic to some finite partition lattice $P_n$. A wealth of other information on subalgebra lattices of Boolean algebras is found in \cite{handbook}. 

For an orthomodular lattice (abbreviated: \textsc{oml}) $L$, we let $\Sub(L)$ be its lattice of subalgebras, and $\BSub(L)$ be its meet-semilattice of Boolean subalgebras. We show that $L$ is determined by $\Sub(L)$ and that $L$ is determined by $\BSub(L)$. We remark that these results use only the order structure of $\Sub(L)$ and $\BSub(L)$, and not the the elements of the subalgebras or the algebraic structure on these elements. This contrasts, for example, with \cite{BergHeunen}. We further prove that if $L,M$ are \textsc{oml}s and $\varphi\colon \Sub(L)\to \Sub(M)$ is a lattice isomorphism, there is an \textsc{oml}-isomorphism $\varphi^*\colon L\to M$ with $\varphi(S)=\varphi^*[S]$ for each subalgebra $S$ of $L$, and such $\varphi^*$ is unique provided $L$ has no 4-element blocks. Similar results hold for an isomorphism $\varphi\colon \BSub(L)\to \BSub(M)$. 
Unfortunately, we do not yet have a result analogous to that of Gr\"{a}tzer {\em et. al.} characterizing the lattices arising as $\Sub(L)$ or the posets arising as $\BSub(L)$ for some \textsc{oml} $L$. 

Apart from their intrinsic interest, these results may shed some additional light on a recent approach to the foundations of quantum mechanics introduced by Isham and Butterfield \cite{Isham1,Isham2}. These authors take as a basic ingredient $\BSub(L)$ for $L$ the \textsc{oml} of projections of a Hilbert space, or more generally, the \textsc{oml} of projections of any von Neumann algebra. That $\BSub(L)$ determines $L$ is perhaps worth noting in this context. 

\pagebreak[3]

This note is arranged in the following manner. The second section provides basic observations about $\BSub(L)$ and shows that a chain-finite \textsc{oml} is determined by the poset $\BSub(L)$. The third section reviews the results of Sachs we use. The fourth section provides our main results. The fifth section discusses some categorical connections, and the sixth provides some directions for further study. The reader should consult \cite{BLT} for general background on lattice theory, and \cite{Kalmbach,Ptak} for background on \textsc{oml}s. 

\section{Basic observations on $\BSub(L)$}

In this section we describe basic properties of the poset of Boolean subalgebras, and show that a chain-finite \textsc{oml} is determined by this poset of Boolean subalgebras. As a comment on notation, we tend to use upper case letters such as $L,M$ for lattices and \textsc{oml}s, lower case letters near the start of the alphabet such as $a,b,c$ for elements of \textsc{oml}s, and lower case letters near the end of the alphabet such as $x,y,z$ for subalgebras of \textsc{oml}s. We do this as subalgebras of $L$ are elements of the poset $\BSub(L)$. Finally, we assume throughout that all Boolean algebras and \textsc{oml}s are non-trivial, meaning that they have at least two elements. 

\begin{defn}
For an \textsc{oml} $L$, let $\BSub(L)$ be the collection of Boolean subalgebras of $L$ partially ordered by set inclusion. 
\end{defn}

We next provide a simple example to illustrate this notion. 

\begin{examp} {\em 
Consider the \textsc{oml} $L$ shown at left below, with $BSub(L)$ at right.

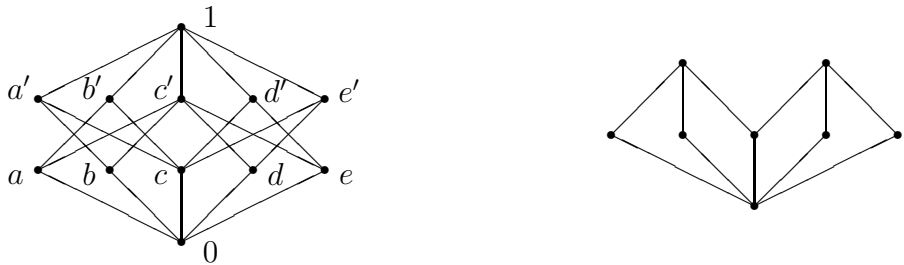
\setlength{\unitlength}{.75in}
\begin{center}
\begin{picture}(6,2.5)(0,-0.5)
\put(1,0){\circle*{0.05}}
\put(0,0.5){\circle*{0.05}}
\put(0.5,0.5){\circle*{0.05}}
\put(1,0.5){\circle*{0.05}}
\put(1.5,0.5){\circle*{0.05}}
\put(2,0.5){\circle*{0.05}}
\put(0,1){\circle*{0.05}}
\put(0.5,1){\circle*{0.05}}
\put(1,1){\circle*{0.05}}
\put(1.5,1){\circle*{0.05}}
\put(2,1){\circle*{0.05}}
\put(1,1.5){\circle*{0.05}}
\put(1,0){\line(-1,1){.5}}
\put(1,0){\line(-2,1){1}}
\put(1,0){\line(0,1){.5}}
\put(1,0){\line(1,1){.5}}
\put(1,0){\line(2,1){1}}
\put(1,1.5){\line(-1,-1){.5}}
\put(1,1.5){\line(-2,-1){1}}
\put(1,1.5){\line(0,-1){.5}}
\put(1,1.5){\line(1,-1){.5}}
\put(1,1.5){\line(2,-1){1}}
\put(1,0.5){\line(-2,1){1}}
\put(1,0.5){\line(-1,1){.5}}
\put(1,0.5){\line(1,1){.5}}
\put(1,0.5){\line(2,1){1}}
\put(1,1){\line(-2,-1){1}}
\put(1,1){\line(-1,-1){.5}}
\put(1,1){\line(1,-1){.5}}
\put(1,1){\line(2,-1){1}}
\put(0,0.5){\line(1,1){.5}}
\put(0,1){\line(1,-1){.5}}
\put(2,0.5){\line(-1,1){.5}}
\put(2,1){\line(-1,-1){.5}}
\put(1.15,0){\makebox(0,0)[tl]{$0$}}
\put(1.15,1.5){\makebox(0,0)[bl]{$1$}}
\put(-0.1,0.5){\makebox(0,0)[tr]{$a$}}
\put(0.4,0.55){\makebox(0,0)[tr]{$b$}}
\put(0.9,0.5){\makebox(0,0)[tr]{$c$}}
\put(1.6,0.55){\makebox(0,0)[tl]{$d$}}
\put(2.1,0.5){\makebox(0,0)[tl]{$e$}}
\put(-0.05,1){\makebox(0,0)[br]{$a'$}}
\put(0.45,1){\makebox(0,0)[br]{$b'$}}
\put(0.95,1){\makebox(0,0)[br]{$c'$}}
\put(1.575,0.975){\makebox(0,0)[bl]{$d'$}}
\put(2.095,0.975){\makebox(0,0)[bl]{$e'$}}
\put(5,0.25){\circle*{0.05}}
\put(4,0.75){\circle*{0.05}}
\put(4.5,0.75){\circle*{0.05}}
\put(5,0.75){\circle*{0.05}}
\put(5.5,0.75){\circle*{0.05}}
\put(6,0.75){\circle*{0.05}}
\put(4.5,1.25){\circle*{0.05}}
\put(5.5,1.25){\circle*{0.05}}
%
\put(5,0.25){\line(-1,1){.5}}
\put(5,0.25){\line(-2,1){1}}
\put(5,0.25){\line(0,1){.5}}
\put(5,0.25){\line(1,1){.5}}
\put(5,0.25){\line(2,1){1}}
%
\put(5,0.75){\line(-1,1){.5}}
\put(5,0.75){\line(1,1){.5}}
%
\put(4,0.75){\line(1,1){.5}}
\put(4.5,0.75){\line(0,1){.5}}
\put(6,0.75){\line(-1,1){.5}}
\put(5.5,0.75){\line(0,1){.5}}

%
\end{picture}
\end{center}
The elements in the poset $BSub(L)$ shown at right are as follows. The bottom element is the subalgebra $\{0,1\}$, and the atoms from left to right are the subalgebras $\{0,1,a,a'\}$, $\{0,1,b,b'\}$, $\{0,1,c,c'\}$, $\{0,1,d,d'\}$, and $\{0,1,e,e'\}$. The maximal element at left is $\{0,1,a,a',b,b',c,c'\}$ and the maximal element at right is $\{0,1,c,c',d,d',e,e'\}$. 
}
\end{examp}

Note, for any \textsc{oml} $L$, each non-empty collection of elements of $\BSub(L)$ has a meet, and each up-directed subset has a join. Further obvious properties are listed below. We recall that, for an \textsc{oml} $L$,  a \emph{block} of~$L$ is a maximal Boolean subalgebra of~$L$. Also, for an element $x\in\BSub(L)$, we write $[0,x]$ for the interval 
$\{y\in\BSub(L): 0\le y\le x\}$ of the poset $\BSub(L)$.

\begin{prop}
Suppose $L$ is an \textsc{oml}. 
\begin{enumerate}
\item The maximal elements of $\BSub(L)$ are the blocks of $L$.
\item Every element of $\BSub(L)$ lies beneath a maximal element of $\BSub(L)$. 
\item The least element of $\BSub(L)$ is the subalgebra $\{0,1\}$ of $L$.
\item The atoms of $\BSub(L)$ are the subalgebras $\{0,a,a',1\}$ where $a\in L-\{0,1\}$. 
\item Each element of $\BSub(L)$ is the join of the atoms beneath it.
\item The interval $[0,x]=\{y\in \BSub(L): 0\le y\le x\}$ of $\BSub(L)$ is the lattice of subalgebras of the Boolean algebra $x$. 
\end{enumerate}
\end{prop}

Clearly a Boolean algebra is finite if and only if it has only finitely many subalgebras. The final condition above allows us to easily recognize the finite Boolean subalgebras of $L$.

\begin{defn}
Say $x\in \BSub(L)$ has finite height if the interval $[0,x]$ is finite, and then let the height of $x$ be one less than the size of a maximal chain in $[0,x]$. 
\end{defn}

So the elements of $\BSub(L)$ of finite height are the finite Boolean subalgebras of $L$; the elements of height 1 are the atoms of $\BSub(L)$ and these correspond to 4-element Boolean subalgebras of $L$; and the elements covering atoms are those of height 2, and these correspond to 8-element Boolean subalgebras of $L$. 
The following small observations are key. 

\begin{lemma}
Let $L$ be an \textsc{oml}, $x=\{0,a,a',1\}$, $y=\{0,b,b',1\}$ for some $a,b\in L-\{0,1\}$. 
\begin{enumerate}
\item $x,y$ are atoms of $\BSub(L)$. 
\item $a,b$ are commuting elements of $L$ iff $x\vee y$ exists in $\BSub(L)$. 
\item $a$ is comparable to one of $b,b'$ iff $x\vee y$ exists and has height at most 2.
\end{enumerate}
\end{lemma}

\noindent {\bf Proof. } The first statement is trivial. The second follows as $a$ commutes with $b$ iff $a,b$ lie in a Boolean subalgebra of $L$. For the third, if $a$ is comparable to one of $b,b'$, then $a$ and $b$ commute, so lie in a Boolean subalgebra of $L$, and the Boolean subalgebra generated by two comparable elements has at most 8 elements. Conversely, if $a,b$ lie in a Boolean subalgebra with at most 8 elements, then $a$ must be comparable to either $b$ or $b'$. $\Box$

\begin{lemma}
Suppose $L$ is an \textsc{oml}, $a\in L-\{0,1\}$ and $x=\{0,a,a',1\}$. The following are equivalent:
\begin{enumerate} 
\item At least one of $a,a'$ is an atom of $L$.
\item For each atom $y$ of $\BSub(L)$, if $x\vee y$ exists then it has height at most 2.
\end{enumerate}
Further, both $a,a'$ are atoms of $L$ iff $x$ is a maximal element of $\BSub(L)$. 
\label{lem:atom}
\end{lemma}

\noindent {\bf Proof. } It is well known~\cite{Kalmbach} that an element of $L$ belonging to a block $z$ is an atom of $L$ iff it is an atom of $z$. To see $(1)\Rightarrow (2)$ suppose either $a$ or $a'$ is an atom of $L$. If $y$ is an atom of $\BSub(L)$, then $y=\{0,b,b',1\}$ for some $b\in L-\{0,1\}$. If $x\vee y$ exists, then $a$ commutes with $b$, so $a,b$ belong to some block $z$ of $L$. Then either $a$ or $a'$ is an atom of $z$, and it follows that $a$ is comparable to either $b$ or $b'$. So $x\vee y$ has height at most 2. To see $(2)\Rightarrow (1)$, find a block $z$ of $L$ containing $a$. Condition~(2) implies for each $b\in z$ that $a$ is comparable to one of $b,b'$, and this implies that one of $a,a'$ is an atom of $z$, hence an atom of $L$. The further comment is trivial. $\Box$
\vspace{2ex}

Already, these simple observations are enough to treat the situation for chain-finite \textsc{oml}s. 

\begin{thm}
If one is given a poset $P$ and told $P$ is isomorphic to $\BSub(L)$ for some finite \textsc{oml} $L$, then one can determine, up to isomorphism, the \textsc{oml} $L$.
\label{thm:chainfinite}
\end{thm}

\noindent {\bf Proof. } Note first that any chain-finite \textsc{oml} $L$ is determined by its atom structure $(A,\perp)$ where $A$ is the set of atoms of $L$ and $\perp$ is the relation of orthogonality on $A$. Indeed, $L$ is isomorphic to the set of subsets $S\subseteq A$ having the property that $S=S^{\perp\perp}$,
where $S^\perp=\{a\in A: a\perp b$ for all $b\in S\}$. 
From $P$, we will construct a structure $(X,\perp)$ isomorphic to $(A,\perp)$, showing that $P$ determines $L$, up to isomorphism. 

Guided by Lemma~\ref{lem:atom}, we let $U$ be the set of atoms of $P$ that satisfy condition~(2) of Lemma~\ref{lem:atom} and are not maximal in $P$, and $V$ be the set of atoms of $P$ that satisfy~(2) and are maximal in $P$. For each $v\in V$ create two elements $v_1,v_2$. Set $X=U\cup\{v_1,v_2:v\in V\}$. We define $\perp$ on $X$ by setting $v_1\perp v_2$ and $v_2\perp v_1$ for each $v\in V$. This reflects that the atoms of a 4-element block are orthogonal to each other and not to any other atoms. For $u,w\in U$ we set $u\perp w$ if $u\neq w$ and $u\vee w$ exists, reflecting that any two distinct atoms belonging to a common block are orthogonal. Then $(X,\perp)$ is isomorphic to $(A,\perp)$ as required. $\Box$
\vspace{2ex}

\section{Results of Sachs}

In this section we review results of \cite{Sachs}. To make the paper self-contained, we include proofs of the key facts. Throughout, for a Boolean algebra $B$, we use $\Sub(B)$ for the collection of subalgebras of $B$ ordered by set inclusion. 

\begin{defn}
A subalgebra of a Boolean algebra $B$ is a dual subalgebra if it consists of an ideal and its complementary filter. It is called a principal dual subalgebra (p.d.\ subalgebra) if it consists of a principal ideal and its complementary filter. 
\end{defn}

Key results of \cite{Sachs} give purely order-theoretic characterizations of the elements of $\Sub(B)$ given by dual, and by p.d.\ subalgebras. We also give such characterizations, but in a way that somewhat simplifies the approach of \cite{Sachs}.

\begin{lemma}
If $x$ is a dual subalgebra of $B$ and $a,b\not\in x$, then the following are equivalent:
\begin{enumerate}
\item $b$ is in the subalgebra generated by $x\cup \{a\}$.
\item $a$ is in the subalgebra generated by $x\cup \{b\}$.
\end{enumerate}
\label{lem:cover}
\end{lemma}

\noindent {\bf Proof. } 
Suppose $x=I\cup F$ where $I$ is a (uniquely determined) proper ideal and $F=\{i':i\in I\}$. If $b$ is in the subalgebra generated by $x\cup \{a\}$, then $b/I$ is in the subalgebra of $B/I$ generated by $a/I$. Then as $b\not\in x$, either $b=a/I$ or $b=a'/I$. Without loss of generality, assume the first. Then $b\vee i = a\vee i$ for some $i\in I$. Then $a=(a\vee i)\wedge (a\vee i') = (b\vee i)\wedge f$ where $f$ is the element $a\vee i'$ of $F$. So $a$ belongs to the subalgebra generated by $x\cup\{b\}$. $\Box$
\vspace{2ex}

We now provide an order-theoretic characterization of dual subalgebras. 

\begin{prop}\label{prop:dsub}
For a subalgebra $x$ of $B$, the following are equivalent:
\begin{enumerate}
\item $x$ is a dual subalgebra. 
\item For all atoms $y$ of $\Sub(B)$ that are incomparable to $x$, we have $x\vee y$ covers $x$. 
\end{enumerate}
\end{prop}

\noindent {\bf Proof. } Suppose $x$ is a dual subalgebra. If $y$ is an atom that is incomparable to $x$, then $y=\{0,a,a',1\}$ for some $a\not\in x$. Clearly $x<x\vee y$. Suppose $z$ is such that $x<z\leq x\vee y$. As $x<z$ there is some $b\in z$ with $b\not\in x$. As $z\leq x\vee y$ we have $b$ is in the subalgebra generated by $x\cup \{a\}$, so by Lemma~\ref{lem:cover} we have $a$ is in the subalgebra generated by $x\cup \{b\}$. So $y\leq z$, and as $x<z$, we have $x\vee y = z$. Thus $x\vee y$ covers $x$.

To show the second statement implies the first, we first observe that as $x$ is a subalgebra of $B$, the set $I=\{a:[0,a]\subseteq x\}$ is an ideal of $B$ and $F=\{a:[a,1]\subseteq x\}=\{i':i\in I\}$ is its dual filter. If $x=I\cup F$, then $x$ is a dual subalgebra. 

Assume $x$ is not a dual subalgebra. We will produce an atom $y$ that is incomparable to $x$ with $x\vee y$ not covering $x$. By the above comments, it follows that $x\neq I\cup F$. So there is some $a\in x$ with $[0,a]\not\subseteq x$ and $[a,1]\not\subseteq x$. Choose $b<a$ and $c<a'$ with $b,c\not\in x$. Set $d=b\vee c$. Note that $a\wedge d = a\wedge (b\vee c) = (a\wedge b)\vee (a\wedge c) = b$ and similarly $a'\wedge d = c$.

Consider the atom $y=\{0,d,d',1\}$. As $a\wedge d = b$, and $a\in x$, $b\not\in x$, it follows that $d\not\in x$. So $x\wedge y=0$, showing that the atom $y$ is incomparable to $x$. It is well known, and easy to verify, that the subalgebra generated by $x\cup\{b\}$ is $\{(e_1\wedge b)\vee (e_2\wedge b'):e_1,e_2\in x\}$. We claim $c$ does not lie in this subalgebra. Indeed, if $c=(e_1\wedge b)\vee (e_2\wedge b')$ for some $e_1,e_2\in x$, then as $c\leq a'\leq b'$ we have $c=c\wedge a' = e_2\wedge b'\wedge a' = e_2\wedge a'$, and this contradicts $c\not\in x$. Since $b=a\wedge d$ and $c=a'\wedge d$, both belong to $x\vee y$. Then as $b\not\in x$, the subalgebra generated by $x\cup\{b\}$ properly contains $x$ and does not contain~$c$. Thus this subalgebra it is properly contained in $x\vee y$ and $x\vee y$ does not cover~$x$. $\Box$
\vspace{2ex}

This leads to the following order-theoretic characterization of p.d.\ subalgebras. 

\begin{prop}\label{prop:pdsub}
A dual subalgebra $x$ is a principal dual subalgebra iff one of the following conditions holds:
\begin{enumerate}
\item $x=0$.
\item $x=B$.
\item $x$ is an atom.
\item There is a dual subalgebra $y$ with $x\wedge y$ an atom that is not a dual subalgebra. 
\end{enumerate}
\end{prop}

\noindent {\bf Proof. } Suppose $x=[0,a]\cup [a',1]$. If $a=0$ then $x=0$. If $a$ is a coatom or $1$, then $x=B$. If $a$ is an atom, then $x=\{0,a,a',1\}$ is an atom. Otherwise $y=[0,a']\cup [a,1]$ is a dual subalgebra with $x\wedge y = \{0,a,a',1\}$ an atom that is not a dual subalgebra. 

For the converse, note $0$ and $B$ are principal dual subalgebras. Any dual subalgebra containing only finitely many elements must be principal, so a dual subalgebra that is an atom must be principal. 
For the remaining case assume $x=I\cup F$ where $I$ is an ideal and $F=\{i':i\in I\}$. Assume also that $y=J\cup G$ where $J$ is an ideal and $G=\{j':j\in J\}$, and that $x\wedge y$ is an atom $\{0,a,a',1\}$ that is not a dual subalgebra. Then neither $a,a'$ is an atom. One of $a,a'$ belongs to $I$, and we may assume this is $a$. Then $a$ cannot belong to $J$ as this would imply some $0<b<a$ belongs to $x\wedge y$. So $a$ belongs to $G$. Then there cannot be an element $c\in I$ that is strictly larger than $a$ as this element would belong to $I\cap G$ and hence to $x\wedge y$. So $I$ is a principal ideal, hence $x$ is a principal dual subalgebra. $\Box$

\begin{rmk}
Sachs gave order-theoretic characterizations of dual and p.d.\ subalegbras in his Theorems~1 and 2 of \cite{Sachs}. His mechanism was somewhat different. He first defined the notion of a dual modular element in a lattice as an element $x$ with $(x\vee y)\wedge z= x\vee (y\wedge z)$ for each $z\geq x$ and $(y\vee x)\wedge w = y\vee (x\wedge w)$ for each $w\geq y$. We note that this is related to, but different from, the usage in \cite{Maeda}, and states that $(x,y)$ and $(y,x)$ are dual modular pairs for each element $y\in L$. In his Theorem~1, Sachs shows that the dual modular elements in the lattice $\Sub(B)$ are exactly the dual subalgebras, and in his Theorem~2 then uses this description to characterize the p.d.\ subalgebras. 
\end{rmk}

With an order-theoretic characterization of the p.d.\ subalgebras, we can show a Boolean algebra $B$ is determined by its lattice of subalgebras. This is the content of the following theorem, effectively proved by Sachs in the proof of his Theorem~4. As there is a small amount of work to extract the result below from his proof, we outline the steps. 


\begin{thm}
Suppose $B$ and $C$ are Boolean algebras and $\varphi\colon \Sub(B) \to \Sub(C)$ is a lattice isomorphism. Then, there is an isomorphism $\varphi^*\colon B\to C$ so that $\varphi(x)=\varphi^*[x]$ for each subalgebra $x$ of $B$. Further, except in the case where $B$ has exactly four elements, the isomorphism $\varphi^*$ with this property is unique. 
\label{thm:Sachs}
\end{thm}

\noindent {\bf Proof. } Let $B\modif $ be the set of elements of $B$ that are not coatoms or $1$, and similarly for $C\modif $. If $b\in B\modif $ then $x=[0,b]\cup [b',1]$ is a p.d.\ subalgebra not equal to the top of $\Sub(B)$. 
The dual subalgebras are characterized lattice-theoretically in Prop.~\ref{prop:dsub}
and the p.d.\ subalgebras among dual subalgebras are characterized in Prop.~\ref{prop:pdsub}.
As $\varphi$ is an isomorphism, $\varphi(x)$ is a p.d.\ subalgebra of $C$ that is not the top of $C$. So there is a unique $c\in C\modif $ with $\varphi(x)=[0,c]\cup[c',1]$. Define $\varphi^*(b)=c$. This defines a map from $B\modif $ to $C\modif $, and as $a_1\leq a_2$ implies $[0,a_1]\cup[a_1',1]$ is contained in $[0,a_2]\cup[a_2',1]$, this map is order-preserving. As $\varphi$ is an isomorphism, it has an inverse $\lambda$, and repeating this process we obtain a map $\lambda^*$ from $C\modif $ to $B\modif $ that is order-preserving. One sees that $\varphi^*$ and $\lambda^*$ are mutually inverse, showing $\varphi^*$ is an isomorphism from the poset $B\modif $ to the poset $C\modif $. 

If $B$ has exactly 4 elements, then $B\modif =\{0\}$, hence $C\modif =\{0\}$, so $C$ has 4 elements, and there is an isomorphism $\varphi^*$ from $B$ to $C$ with $\varphi^*[x]=\varphi(x)$ for each subalgebra $x$ of $B$. In fact, there are two such isomorphisms as there are two choices how to map the two atoms of $B$ to the two atoms of $C$. 
Similar arguments lead to a unique isomorphism in the case when $B$ has exactly 2 elements (then $B\modif =C\modif =\emptyset$).

Suppose $B$ has more than 4 elements. Then for each $b\in B-B\modif $ we have $b'\in B\modif$. So we extend $\varphi^*$ to $B$ by setting $\varphi^*(b)=\varphi^*(b')'$ for each $b\in B-B\modif $. This extension is seen to be order-preserving by noting that $d\leq b$ implies $d\wedge b'=0$, hence $\varphi^*(d)\wedge\varphi^*(b')=0$, so $\varphi^*(d)\leq\varphi^*(b')' = \varphi^*(b)$. Similarly $\lambda^*$ extends to $C$, and these extensions $\varphi^*$ and $\lambda^*$ are easily seen to be mutually inverse order-preserving maps between $B$ and $C$, hence Boolean algebra isomorphisms. 

We next show $\varphi^*[x]=\varphi(x)$ for each subalgebra $x$ of $B$. By construction, this is true for each p.d.\ subalgebra $x$. Suppose $b\in B$, let $x=[0,b]\cup[b',1]$ and $y=[0,b']\cup[b,1]$, and note $x\wedge y =\{0,b,b',1\}$. Then as $\varphi^*$ is an isomorphism, $\varphi^*[\{0,b,b',1\}]=\varphi^*[x\wedge y] = \varphi^*[x]\wedge\varphi^*[y] = \varphi(x)\wedge\varphi(y)=\varphi(x\wedge y)=\varphi(\{0,b,b',1\})$. For any subalgebra $z$ we have $z=\bigcup\{\{0,b,b',1\}:b\in z\}$. As $\varphi$ is an isomorphism, $\varphi(z)=\bigcup\{\varphi(\{0,b,b',1\}):b\in z\}=\bigcup\{\varphi^*[\{0,b,b',1\}]:b\in z\} = \varphi^*[\bigcup\{\{0,b,b',1\}:b\in z\}]=\varphi^*[z]$. Finally, any isomorphism from $B$ to $C$ with this property must, by construction, agree with $\varphi^*$, giving uniqueness when $B$ has other than 4 elements. $\Box$

\section{Main results}

\begin{thm}
If $L$ and $M$ are \textsc{oml}s and $\varphi\colon \BSub(L)\to \BSub(M)$ is an isomorphism of posets, then there is an isomorphism $\varphi^*\colon L\to M$ with $\varphi(x)=\varphi^*[x]$ for each Boolean subalgebra $x$ of $L$. Further, the map $\varphi^*$ with this property is unique provided $L$ has no blocks with four elements. 
\label{thm:main}
\end{thm}

\noindent {\bf Proof. } Note that blocks of $L$ are maximal elements of $\BSub(L)$. So a block of $L$ having exactly four elements is simply a maximal element of $L$ that is an atom of $\BSub(L)$. As there is an isomorphism between $\BSub(L)$ and $\BSub(M)$, then $L$ has a block with four elements iff $M$ has one. We assume for now that $L$ has no four-element block. 

For each $x\in \BSub(L)$ we use $\varphi_x$ for the restriction $\varphi|[0,x]$ of $\varphi$ to the interval $[0,x]$ of $\BSub(L)$. As $\varphi$ is an isomorphism, we have $\varphi_x\colon [0,x]\to[0,\varphi(x)]$ is an isomorphism. Note $[0,x]$ is literally equal to the lattice of subalgebras of the Boolean subalgebra $x\subseteq L$, and $[0,\varphi(x)]$ is literally equal to the lattice of subalgebras of the Boolean subalgebra $\varphi(x)\subseteq M$. Then as long as $x$ has more than four elements, in other words as long as $x$ is not an atom of $\BSub(L)$, the above result shows there is a unique Boolean algebra isomorphism $\varphi_x^*\colon x\to \varphi(x)$ with $\varphi_x(y)=\varphi_x^*[y]$ for each subalgebra $y$ of $x$.

We next consider how these maps $\varphi_x^*$ match up. Suppose $x,y$ belong to $\BSub(L)$ and $z=x\wedge y$. This means $z$ is the intersection of the Boolean subalgebras $x,y$ of $L$. As $\varphi$ is an isomorphism, $\varphi(z)$ is the intersection of the Boolean subalgebras $\varphi(x),\varphi(y)$ of $M$. We then have that $\varphi^*_x$ is an isomorphism from $x$ to $\varphi(x)$, $\varphi^*_y$ is an isomorphism from $y$ to $\varphi(y)$, and $\varphi^*_z$ is an isomorphism from $z$ to $\varphi(z)$. It is obvious from the definitions that $\varphi_x$ and $\varphi_y$ agree on $[0,z]$, our aim is to show $\varphi^*_x$ and $\varphi^*_y$ agree on (the elements of)~$z$. 
\vspace{2ex}

\noindent {\bf Claim 1 } Suppose $x,y$ are Boolean subalgebras of $L$ and $z=x\wedge y$ has other than four elements. Then $\varphi^*_x$ and $\varphi^*_y$ agree on $z$. 
\vspace{2ex}

\noindent {\bf Proof. } Note $\varphi^*_x|z$ is an isomorphism from $z$ to $\varphi^*(z)$. Indeed, as $z\leq x$ we have $\varphi^*_x[z]=\varphi_x(z)=\varphi(z)$. Then as $\varphi^*_x$ is an isomorphism from $x$ to $\varphi(x)$, we have $\varphi^*|z$ is an isomorphism from $z$ to $\varphi(z)$. Further, for $w\leq z$ we have $w\leq x$. So $(\varphi^*_x|z)[w]=\varphi^*_x[w]=\varphi_x(w)=\varphi_z(w)$. But $z$ has other than 4 elements, so there is a unique isomorphism $\psi\colon z\to \varphi(z)$ with $\psi[w]=\varphi_z(w)$ for all $w\leq z$. We have shown $\varphi^*_x|z$ is such a map, and by definition $\varphi^*_z$ is another. So $\varphi^*_x|z = \varphi^*_z$. We may apply this argument to show $\varphi^*_y|z$ is also equal to $\varphi^*_z$, so $\varphi^*_x|z=\varphi^*_y|z$, showing $\varphi^*_x$ and $\varphi^*_y$ agree on $z$. $\Box$
\vspace{2ex}

%
\vspace{2ex}

\noindent {\bf Claim 2 } Suppose $x,y$ are blocks of $L$, and $z=x\wedge y$ equals $\{0,b,b',1\}$ where $b$ is an atom of $L$. Then $\varphi^*_x$ and $\varphi^*_y$ agree on $z$. 
\vspace{2ex}

\noindent {\bf Proof. } As $x$ is maximal in $\BSub(L)$ and $\varphi$ is an isomorphism, $\varphi(x)$ is maximal in $\BSub(M)$, so $\varphi(x)$ is a block of $M$. Similarly $\varphi(y)$ is a block of $M$. As $x,y$ are blocks of $L$, neither is contained in the other unless they are equal, and in this case it is clear that $\varphi_x^*$ and $\varphi^*_y$ agree. So both properly contain their intersection, hence both have at least 8 elements. 

As $b$ is an atom of $L$, it is an atom of both $x,y$, and its complement $b'$ is not an atom of either $x,y$ as these blocks have at least 8 elements. As $\varphi^*_x$ is an isomorphism from the block $x$ of $L$ to the block $\varphi(x)$ of $M$, the atom $b$ of $x$ is mapped by $\varphi^*_x$ to an atom of $\varphi(x)$, and the non-atom $b'$ of $x$ is mapped by $\varphi^*_x$ to a non-atom of $\varphi(x)$. As $\varphi(x)$ is a block of $M$, we have $\varphi^*_x(b)$ is an atom of $M$ and $\varphi^*_x(b')$ is a non-atom of $M$. The same argument shows $\varphi^*_y(b)$ is an atom of $M$ and $\varphi^*_y(b')$ is a non-atom of $M$. Now $z=x\wedge y$ is an atom of $\BSub(L)$ and $\varphi$ is an isomorphism, so $\varphi(z)$ is an atom of $\BSub(M)$, hence a four-element subalgebra $\{0,c,c',1\}$ of $M$ that is equal to the intersection of the blocks $\varphi(x)\cap\varphi(y)$. We know $\varphi^*_x[z]=\varphi_x(z)=\{0,c,c',1\}$ and $\varphi^*_y[z]=\varphi_y(z)=\{0,c,c',1\}$. So $\{\varphi^*_x(b),\varphi^*_x(b')\}$ must equal $\{c,c'\}$ and similarly $\{\varphi^*_y(b),\varphi^*_y(b')\}$ also equals $\{c,c'\}$. As $\varphi^*_x(b)$ is an atom of $M$ and $\varphi^*_x(b')$ is not, exactly one of $c,c'$ is an atom of $M$, say this is $c$. Then we have $\varphi^*_x(b)=c$ and $\varphi^*_y(b)=c$. It follows that $\varphi^*_x$ and $\varphi^*_y$ agree on all of $z=\{0,b,b',1\}$. $\Box$
\vspace{2ex}

\noindent {\bf Claim 3 } Suppose $x,y$ are blocks of $L$ and $z=x\wedge y$ equals $\{0,b,b',1\}$ where neither $b,b'$ is an atom of $L$. Then $\varphi^*_x$ and $\varphi^*_y$ agree on $z$. 
\vspace{2ex}

\noindent {\bf Proof. } As $b\in x$ and $b$ is not an atom of $L$, then $b$ is not an atom of the block $x$. So there is $a\in x$ with $0<a<b$. Similarly as $b'\in y$ and $b'$ is not an atom of $L$, then $b'$ is not an atom of $y$, so there is $c\in y$ with $0<c<b'$. Then $a<b<c'$ showing $a,b,c$ belong to a Boolean subalgebra of $L$. Using $\langle \cdot \rangle$ for the subalgebra generated by some set, we let $u=\langle a,b\rangle$, $v=\langle b,c\rangle$, and $w=\langle a,b,c\rangle$. Then as $u,v$ each have a non-trivial comparability, they have at least 8 elements, hence so also does $w$. We note $z < u < x$, and $z<v<y$. 

Apply Claim~1 to $x,u$ noting $u=x\wedge u$ has more than $4$ elements. We then obtain $\varphi^*_x$ and $\varphi^*_u$ agree on $u$. In particular, $\varphi^*_x(b)=\varphi^*_u(b)$. Similarly, as $v=y\wedge v$ has more than 4 elements, we have $\varphi^*_v(b)=\varphi^*_y(b)$. We next apply Claim~1 to $w,u$ noting $u=w\wedge u$ has more than four elements, to obtain $\varphi^*_u(b)=\varphi^*_w(b)$, and similarly as $v=w\wedge v$ has more than 4 elements, we obtain $\varphi^*_w(b)=\varphi^*_v(b)$. So $\varphi^*_x(b)=\varphi^*_u(b)=\varphi^*_w(b)=\varphi^*_v(b)=\varphi^*_y(b)$. It follows that $\varphi^*_x$ and $\varphi^*_y$ agree on all of $z=\{0,1,b,b'\}$. $\Box$
\vspace{2ex}

We have shown the following. 
\vspace{2ex}

\noindent {\bf Claim 4 } If $x,y$ are blocks of $L$, then $\varphi^*_x$ and $\varphi^*_y$ agree on $z=x\wedge y$. 
\vspace{2ex}

We may therefore define 
\[\varphi^*\colon L\to M \quad\mbox{by setting $\varphi^*(b)=\varphi^*_x(b)$ if $b$ belongs to the block $x$}.\]
This is a well-defined map as if $b$ belongs to the blocks $x$ and $y$, then $\varphi^*_x(b)$ and $\varphi^*_y(b)$ agree. 
\vspace{2ex}

\noindent {\bf Claim 5 } For any $y\in \BSub(L)$ we have $\varphi(y)=\varphi^*[y]$. 
\vspace{2ex}

\noindent {\bf Proof. } Indeed, such $y$ is contained in some block $x$ of $L$. Then by definition, $\varphi^*[y] = \varphi_x^*[y]$. But $\varphi^*_x$ was chosen to have the property that $\varphi^*[y]=\varphi_x(y)$ for all $y\leq x$. Thus $\varphi^*[y]=\varphi_x(y)$, and as $\varphi_x(y)$ equals $\varphi(y)$, we have $\varphi(y)=\varphi^*[y]$. $\Box$
\vspace{2ex}

\noindent {\bf Claim 6 } $\varphi^*$ is an \textsc{oml}-isomorphism. 
\vspace{2ex}

\noindent {\bf Proof. } Note $\varphi^*$ preserves $0,1$ since $\varphi^*_x$ is a Boolean algebra homomorphism for each block~$x$. Also $\varphi^*$ preserves orthocomplementation as $b\in x$ implies $b'\in x$ and $\varphi_x^*(b')=\varphi_x^*(b)'$. We note also that $\varphi^*$ is order-preserving. Indeed, if $a\leq b$, then there is a block $x$ with $a,b\in x$. Then $\varphi^*(a)=\varphi^*_x(a)\leq\varphi^*_x(b)=\varphi^*(b)$ where the partial ordering in this expression is that in the block $\varphi(x)$, and this implies $\varphi^*(a)\leq\varphi^*(b)$ in $M$. 

As $\varphi\colon \BSub(L)\to \BSub(M)$ is an isomorphism of posets, there is an inverse map \linebreak[4] $\lambda\colon \BSub(M)\to \BSub(L)$. We can then define $\lambda^*\colon M\to L$ by the exact process as above. We will then have that $\lambda^*$ is order-preserving. If we can show that $\varphi^*$ and $\lambda^*$ are mutually inverse, then it will follow that $\varphi$ is an order isomorphism, hence a lattice isomorphism, hence an ortholattice isomorphism. 

Let $b\in L$ belong to the block $x$ of $L$. Then for the restrictions $\varphi_x\colon [0,x]\to[0,\varphi(x)]$ and $\lambda_{\varphi(x)}\colon [0,\varphi(x)]\to[0,x]$, we have these are mutually inverse isomorphisms between the lattices of subalgebras of $x$ and $\varphi(x)$. As we have assumed all blocks have at least 8 elements, there are unique isomorphisms $\varphi_x^*\colon x\to\varphi(x)$ and $\lambda_{\varphi(x)}^*\colon \varphi(x)\to x$ with $\varphi_x^*[y]=\varphi_x(y)$ for all $y\leq x$ and $\lambda_{\varphi(x)}^*[p]=\lambda_{\varphi(x)}(p)$ for all $p\leq \varphi(x)$. From considerations in the Boolean case, or by applying uniqueness to the composite and the identity map, we have $\varphi_x^*$ and $\lambda_{\varphi(x)}^*$ are mutually inverse isomorphisms. So $\varphi^*$ and $\lambda^*$ are mutually inverse, as required. $\Box$
\vspace{2ex}

\noindent {\bf Claim 7 } If $\psi\colon L\to M$ satisfies $\psi[y]=\varphi(y)$ for each $y\in \BSub(L)$, then $\psi=\varphi^*$. 
\vspace{2ex}

\noindent {\bf Proof. } Suppose $b\in L$ and $b$ belongs to the block $x$ of $L$. As $\psi[x]=\varphi(x)$ we have that the restriction $\psi|x$ of $\psi$ to $x$ is an isomorphism from $x$ to $\varphi(x)$. Further, for each $y\leq x$ we have $(\psi|x)[y]=\psi[y]=\varphi(y)=\varphi_x(y)$. But $\varphi^*_x$ is the unique isomorphism from $x$ to $\varphi(x)$ with this property, so $\psi|x=\varphi^*_x$. As $b\in x$ we have $\psi(b)=(\psi|x)(b)=\varphi^*_x(b)=\varphi^*(b)$. So $\psi=\varphi^*$. $\Box$
\vspace{2ex}

We have proved our result in the case that $L$ has no blocks with four elements. Suppose $L, M$ are arbitrary with $\varphi\colon \BSub(L)\to \BSub(M)$ a poset isomorphism. Let $L\modif $ be the \textsc{oml} obtained from $L$ by removing all elements that are both atoms and coatoms, and let $M\modif $ be obtained similarly from~$M$. As 4-element blocks of $L$ are exactly those atoms of $\BSub(L)$ that are maximal, $\varphi$ provides a bijection between the 4-element blocks of $L$ and those of $M$, so they have the same number of 4-element blocks. Enumerate the 4-element blocks of $L$ as $z_\alpha$ $(\alpha\in\kappa)$, so the 4-element blocks of $M$ are exactly the $\varphi(z_\alpha)$ $(\alpha\in\kappa)$. Then $\BSub(L\modif )$ is simply $\BSub(L)$ with the elements $z_\alpha$ $(\alpha\in\kappa)$ removed, and $\BSub(M\modif )$ is simply $\BSub(M)$ with the elements $\varphi(z_\alpha)$ $(\alpha\in\kappa)$ removed. So $\varphi$ restricts to a poset isomorphism $\varphi\modif $ between $\BSub(L\modif )$ and $\BSub(M\modif )$. As $L\modif ,M\modif $ are \textsc{oml}s with no 4-element blocks, we get an isomorphism $(\varphi\modif )^*\colon L\modif \to M\modif $ with $(\varphi\modif )^*[y]=\varphi(y)$ for every subalgebra $y$ of $L\modif $. We extend $(\varphi\modif )^*$ to a map $\varphi^*$ from $L$ to $M$ by choosing for each $\alpha\in\kappa$ one of the two possible isomorphisms from the 4-element block $z_\alpha$ of $L$ to the block $\varphi(z_\alpha)$ of $M$. It is clear that $(\varphi\modif )^*$ is an isomorphism from $L\modif $ to $M\modif $ with $(\varphi\modif )^*[y]=\varphi\modif (y)$ for each $y\in \BSub(L\modif )$. Therefore $\varphi^*$ is an isomorphism from $L$ to $M$ with $\varphi^*[y]=\varphi(y)$ for each $y\in \BSub(L)$. $\Box$
\vspace{2ex}

Having established Theorem~\ref{thm:main}, we find consequences of this result
for the \emph{lattices} of \emph{all} (not only Boolean) subalgebras of \textsc{oml}s. 

\begin{thm}
If $L,M$ are \textsc{oml}s and $\varphi\colon \Sub(L)\to \Sub(M)$ is a lattice isomorphism, then there is an isomorphism $\varphi^*\colon L\to M$ with $\varphi^*[x]=\varphi(x)$ for each subalgebra $x$ of $L$. Further, if $L$ has no 4-element blocks, this map $\varphi^*$ is unique. 
\end{thm}

\noindent {\bf Proof. } We first show that if $K$ is an \textsc{oml} and $\Sub(K)$ is isomorphic to $\Sub(B)$ for some Boolean algebra $B$, then $K$ is Boolean. Indeed, if $K$ is not Boolean, there are two non-commuting elements in $K$, hence a non-Boolean two-generated subalgebra of $K$. As the free \textsc{oml} on two generators is $\MO_2\times 2^4$, where $\MO_2$ is the modular ortholattice with $0,1$ and four atoms \cite{Kalmbach}, there is a non-Boolean quotient of this that is a subalgebra of $K$. One can check that this implies $K$ has a subalgebra $x$ isomorphic to either $\MO_2$ or $\MO_2\times 2$. This means the interval $[0,x]$ of $\Sub(K)$ is isomorphic to $\Sub(\MO_2)$ or $\Sub(\MO_2\times 2)$. But $\Sub(\MO_2)$ has 2 atoms and $\Sub(\MO_2\times 2)$ has 5 atoms, so neither can be an interval of $\Sub(B)$ for a Boolean algebra $B$ as the number of atoms in the subalgebra lattice of a Boolean algebra with $2^n$ elements is $2^{n-1}-1$. 

So $\BSub(L)$ consists exactly of the elements $x\in \Sub(L)$ with $[0,x]$ isomorphic to $\Sub(B)$ for some Boolean algebra $B$. As $\varphi$ is an isomorphism from $\Sub(L)$ to $\Sub(M)$, we have that $\varphi$ restricts to an isomorphism from $\BSub(L)$ to $\BSub(M)$. Then by Theorem~\ref{thm:main} there is an isomorphism $\varphi^*\colon L\to M$ with $\varphi^*[y]=\varphi(y)$ for each Boolean subalgebra $y$ of $L$, and if $L$ has no 4-element blocks this is unique. For each $a\in L$ the subalgebra $\{0,a,a',1\}$ is Boolean, so $\varphi^*[\{0,a,a',1\}]=\varphi(\{0,a,a',1\})$. As every subalgebra of $L$ is the union of ones of this form, as in the proof of Theorem~\ref{thm:Sachs}, this shows $\varphi^*[x]=\varphi(x)$ for each subalgebra $x$ of $L$. $\Box$

\begin{coro}
An \textsc{oml} $L$ is determined up to isomorphism by the poset $\BSub(L)$, and is determined up to isomorphism by the lattice $\Sub(L)$. Therefore each of $\Sub(L)$ and $\BSub(L)$ determines, up to isomorphism, the other. 
\end{coro}

\begin{rmk}
These results can in part be extended to orthomodular posets (abbreviated: \textsc{omp}s). For an \textsc{omp} $P$, we say \cite{Ptak,regular} a subset $B\subseteq P$ is a Boolean subalgebra of $P$ if $B$ consists of pairwise commuting elements and is closed under orthocomplementation and the necessarily existing finite joins and finite meets of $P$ (commuting elements in an \textsc{omp} always have a join and meet). One then defines $\BSub(P)$ in the obvious manner, and notes that if $P$ is in fact an \textsc{oml}, then this definition of $\BSub(P)$ agrees with the earlier one. 
The proof of Theorem~\ref{thm:main} is valid, word for word, in this more general setting. 
\end{rmk}


\begin{rmk}
One might ask whether these results extend also to the setting of ortholattices (abbreviated: \textsc{ol}s). This is not the case. The underlying reason is that the orderings of the Boolean subalgebras of an \textsc{oml} determine the ordering of an \textsc{oml}, but this does not apply to \textsc{ol}s. For a specific example, consider the 6-element non-modular \textsc{ol} $L$ commonly referred to as the benzene ring~\cite{Beran}. For it, we have $\BSub(L)$ consists of a bottom element and two atoms covering this, and $\Sub(L)$ is simply formed by adding a top element to $\BSub(L)$. These structures are isomorphic to $\BSub(\MO_2)$ and $\Sub(\MO_2)$ respectively.
\end{rmk}

\section{Categorical implications}

Let $\Bbb{OML}$ be the category with \textsc{oml}s as objects and \textsc{oml}-homomorphisms as morphisms, and let $\Bbb{ALG}$ be the category with algebraic lattices as objects and maps that preserve arbitrary meets and non-empty up-directed joins as morphisms. As with any variety of algebras considered as a category, we trivially have the following. 

\begin{prop}
There is a contravariant functor $Sub\colon \Bbb{OML}\to \Bbb{ALG}$ taking an \textsc{oml} $L$ to $\Sub(L)$, and an \textsc{oml}-homomorphism $f\colon L\to M$ to the map $f^{-1}\colon \Sub(M)\to \Sub(L)$. 
\end{prop}

Recall \cite{Herrlich} that a contravariant functor $F\colon \Bbb{C}\to\Bbb{D}$ is \emph{dense} if each object in $\Bbb{D}$ is isomorphic to one in the image of $F$; \emph{full} if the restriction of $F$ to each homset $\Bbb{C}(A,B)$ is onto $\Bbb{D}(FB,FA)$; and \emph{faithful} if the restriction of $F$ to each homset is one-one. We say $F$ is a \emph{dual equivalence} if it is dense, full, and faithful. 

\begin{prop}
$Sub\colon \Bbb{OML}\to\Bbb{ALG}$ is not dense, full, or faithful. 
\end{prop}

\noindent {\bf Proof. } It is easy to find an algebraic lattice that is not equal to $\Sub(L)$ for any \textsc{oml} $L$, for instance the 3-element chain, so the functor $Sub$ is not dense. The 4-element Boolean algebra $B$ has a non-trivial automorphism $\alpha$, and $\alpha^{-1}$ and $id^{-1}$ are both the identity map on $\Sub(B)$, so $Sub$ is not faithful. Finally, consider the 8-element Boolean algebra $B$ with atoms $a,b,c$. The subalgebra $\Sub(B)$ has a bottom, top, and three atoms. Consider the map $\varphi\colon \Sub(B)\to \Sub(B)$ that maps one of these atoms, say $\{0,c,c',1\}$, to the bottom, and leaves all other elements fixed. One can see that $\varphi$ preserves all meets, and as $\Sub(B)$ is finite, trivially preserves non-empty up-directed joins. But $\varphi$ cannot equal $f^{-1}$ for any homomorphism $f\colon B\to B$ as $\varphi(\{0,1\})=\{0,1\}$ would imply $f$ is one-one, and $\varphi(\{0,c,c',1\})=\{0,1\}$ would imply $f$ is not onto as nothing is mapped to $c$. So $Sub$ is not full. $\Box$
\vspace{2ex}

For any contravariant functor $F\colon \Bbb{C}\to\Bbb{D}$, by restricting to suitable subcategory $\Bbb{D}\modif $ we obtain a functor $F\modif \colon \Bbb{C}\to \Bbb{D}\modif $ that is dense and full. In the following, we consider more closely the matter of faithfulness. Here we let $\Image f$ be the image of $f$. 

\begin{prop}
Suppose $L,M$ are \textsc{oml}s and $f\colon L\to M$. 
\begin{enumerate}
\item If $\Image f$ has two elements, then any $g\colon L\to M$ with $\Image f = \Image g$ has $f^{-1}=g^{-1}$. 
\item If $\Image f$ has a 4-element block, there is a $g\colon L\to M$ with $f^{-1}=g^{-1}$ and $f\neq g$. 
\item Otherwise, if $g\colon L\to M$ and $f^{-1}=g^{-1}$, then $f=g$. 
\end{enumerate}
In the first case above, there may, or may not, be such a $g$ that is different from $f$. 
\end{prop}

\noindent {\bf Proof. } For the first statement, suppose $f,g\colon L\to M$ both have image $\{0,1\}$. Then $f^{-1}[x]=L$ and $g^{-1}[x]=L$ for every subalgebra $x$ of $M$, so $f^{-1}=g^{-1}$. To see the further remark regarding the first statement, note that a Boolean algebra with more than 2 elements will have different maps onto a 2-element Boolean algebra. On the other hand, the \textsc{oml} of finite and cofinite dimensional subspaces of a Hilbert space has only a single map onto the 2-element Boolean algebra. For the second statement, define an automorphism $\alpha$ of $M$ to be the non-trivial automorphism on the 4-element block of $M$ which is the image of $f$ and is the identity elsewhere. Then $f$ and $\alpha\circ f$ are distinct, but $f^{-1}$ and $(\alpha\circ f)^{-1}$ agree. 

We consider the third statement. Suppose $f,g\colon L\to M$ with $f^{-1}=g^{-1}$. As the image of $f$ is the least element of $\Sub(M)$ mapped by $f^{-1}$ to the top of $\Sub(L)$, it follows that the image of $f$ equals that of $g$. 
\vspace{2ex}

\noindent {\bf Claim 1 } For each $a\in L$, $g(a)$ equals either $f(a)$ or $f(a)'$. 
\vspace{2ex}

\noindent {\bf Proof. } As $f^{-1}[\{0,1\}] = g^{-1}[\{0,1\}]$, we have that $f(a)$ belongs to $\{0,1\}$ iff $g(a)$ belongs to $\{0,1\}$. In particular, our claim is established if $f(a)$ is either $0,1$. Suppose $f(a)$ does not equal $0,1$, so $g(a)$ also does not equal $0,1$. As $a\in f^{-1}[\{0,f(a),f(a)',1\}] = g^{-1}[\{0,f(a),f(a)',1\}]$, it follows that $g(a)$ is either $f(a)$ or $f(a)'$. $\Box$
\vspace{2ex}

\noindent {\bf Claim 2 } For each $a\in L$, $f(a)=0$ iff $g(a)=0$.
\vspace{2ex}

\noindent {\bf Proof. } From Claim 1, the only way this can fail to happen is if there is some $a\in L$ with $f(a)=0$ and $g(a)=1$. As the image of $f$ has more than 2 elements, there is some $m$ in this image with $0<m<1$. Choose some $b\in L$ with $f(b)=m$. Then $f(a\vee b)=0\vee m = m$. As $a\vee b\geq a$ we have $g(a\vee b)\geq g(a) = 1$, showing $g(a\vee b)=1$. This contradicts Claim~1 as $g(a\vee b)$ equals neither $f(a\vee b)$ nor $f(a\vee b)'$. $\Box$
\vspace{2ex}

Finally, we show that $f(a)=g(a)$ for any $a\in L$. From Claim~2, we may assume $f(a)\neq 0$, and using the fact that $f,g$ are homomorphisms, we may assume $f(a)\neq 1$. So assume there is some $a\in L$ with $f(a)\neq 0,1$ and $g(a)\neq f(a)$. Then by Claim~1 $g(a)=f(a)'$. Suppose $f(a)=m$. Then $0<m<1$ and $g(a)=m'$. As the image of $f$, which equals the image of $g$, has no 4-element blocks, there is some element $n$ in this image with either $0<n<m<1$ or $0<n<m'<1$. As we may work with $f(a'),g(a')$ instead of $f(a),g(a)$, we may assume $0<n<m<1$. As $n$ belongs to the image of $f$ there is some $b\in L$ with $f(b)=n$. Considering $a\wedge b$ if necessary, we may assume $b<a$. By Claim~1, $g(b)$ must be either $f(b)=n$ or $f(b)'=n'$. But $b\leq a$ implies $g(b)\leq g(a)=m'$, and as $n<m$, neither $n,n'$ is less than $m'$. This provides a contradiction. So $g(a)=f(a)$. $\Box$

\begin{defn}
Let $\Bbb{OML}\modif $ be the category whose objects are \textsc{oml}s having no blocks with 4 or fewer elements and whose morphisms are onto \textsc{oml}-homomorphisms. 
\end{defn}

In this category any $f\colon L\to M$ satisfies condition~(3) of the above proposition. This provides the following. 

\begin{coro}
The functor $Sub\colon \Bbb{OML}\modif \to\Bbb{ALG}$ is faithful, and therefore induces a dual equivalence between $\Bbb{OML}\modif $ and a subcategory of $\Bbb{ALG}$. 
\end{coro}

It is not clear that these categorical considerations are more than a curiosity. Still, they seem worth the small investment of effort past our main results. 

\section{Conclusions}

To conclude, we mention several problems related to this work. 

\begin{problem}
Suppose $L$ is an \textsc{oml} and $P$ is a poset isomorphic to $\BSub(L)$. Find a direct construction of an \textsc{oml} isomorphic to $L$ from the poset $P$. Similarly, find a way to directly construct an \textsc{oml} isomorphic to $L$ from a lattice isomorphic to $\Sub(L)$. 
\end{problem}

This first problem may be related to what is known as ``pasting'' families of Boolean algebras \cite{Kalmbach}. Alternately, the method of Gr\"{a}tzer \textit{et al.}\ \cite{Gratzer} of constructing a Boolean algebra from $\Sub(B)$ via a direct limit of finite Boolean algebras may perhaps be of use. 

\begin{problem}
Characterize the posets that arise as $\BSub(L)$ and the lattices that arise as $\Sub(L)$ for some \textsc{oml} $L$. 
\end{problem}

This second problem may be related to results involving ``loops'' in \textsc{oml}s \cite{Kalmbach}. 

\begin{problem}
For an \textsc{oml} $L$, each of $\BSub(L)$ and $\Sub(L)$ determines the other. We have provided a simple order-theoretic way to construct a $\BSub(L)$ from $\Sub(L)$. Provide a direct order-theoretic construction of $\Sub(L)$ from $\BSub(L)$. 
\end{problem}

\begin{problem} 
Characterize order-theoretically the maps that arise as $f^{-1}\colon \Sub(M)\to \Sub(L)$ for some homomorphism $f\colon L\to M$. 
\end{problem}

Finally, the following is related to work initiated by Isham and Butterfield \cite{Isham1,Isham2}.

\begin{problem}
To what extent is a $C^*$-algebra determined by its poset of abelian $C^*$-subalgebras?
\end{problem}

We make a few comments on this last question. An abelian C*-algebra is determined up to *-isomorphism by its lattice of C*-subalgebras \cite[Theorem~11]{Mendivil}. In the non-abelian case, we note that a C*-algebra and its opposite have exactly the same subalgebras. So by results in \cite{Connes} we cannot in general hope to recover a C*-algebra $A$ from its poset of abelian C*-subalgebras. Perhaps recovering $A$ up to its Jordan structure might be a reasonable goal. In the setting of von Neumann algebras, this is what is achieved in \cite{DoeringHarding}. Finally, we thank C.~Heunen for pointing out reference \cite{Connes} to us.


\end{document}